\documentclass[doublecol]{epl2} 
\usepackage{epsfig} 
\usepackage{graphics}
\usepackage{xspace}
\usepackage{amsmath}
\usepackage{booktabs}
\usepackage{color}
\usepackage{subfigure}

\newcommand{\et}{{et al.}\xspace}

\definecolor{Red}{rgb}{0.7,0.0,0.0}
\definecolor{Green}{rgb}{0.0,0.7,0.0}
\definecolor{Blue}{rgb}{0.0,0.0,0.7}

\title{Towards Evidence of Long-Range Correlations in Shallow Seismic Activities}
\shorttitle{Towards Evidence of Long-Range Correlations in Shallow Seismic Activities} %Insert here a short version of the title if it exceeds 70 characters

\author{Douglas Ferreira\inst{1} \and Jennifer Ribeiro\inst{1} \and Andr\'es Papa\inst{2,3} \and Ronaldo Menezes\inst{4}}
\shortauthor{Douglas Ferreira \etal}

\institute{                    
  \inst{1} LISComp Laboratory, Instituto Federal do Rio de Janeiro - Paracambi, RJ, Brazil \\
  \inst{2} Observat\'orio Nacional, Rio de Janeiro, RJ, Brazil \\
  \inst{3} Instituto de F\'isica, Universidade do Estado do Rio de Janeiro, Rio de Janeiro, RJ, Brazil \\
  \inst{4} BioComplex Laboratory, School of Computing, Florida Institute of Technology, Melbourne, USA
}

\pacs{89.75.Hc}{Networks and genealogical trees}
\pacs{91.30.Px}{Earthquakes}
\pacs{05.90.+m}{Other topics in statistical physics, thermodynamics, and nonlinear dynamical systems }

\abstract{
In this work, we introduce a new methodology to construct a network of epicenters that avoids problems found in well-established methodologies when they are applied to global catalogs of earthquakes located in shallow zones. The new methodology involves essentially the introduction of a time window which works as a temporal filter. Our approach is more generic and for small regions the results coincide with previous findings. The network constructed with that model has small-world properties and the distribution of node connectivity follows a non-traditional function, namely a $q$-exponential, where scale-free properties are present. The vertices with larger connectivity in the network correspond to the areas with very intense seismic activities in the period considered. These new results strengthen the hypothesis of long spatial and temporal correlations between earthquakes.}

\begin{document}

\maketitle

\section{Introduction}
\label{introduction}

Amongst many natural disasters observed in nature, earthquakes are one of the most devastating not only by the number of lives lost but also by the economic damage they cause (e.g. the 1976 China earthquake is estimated to have killed more than 230,000 people). The understanding of seismic phenomena is of crucial importance in engineering, and in social, geophysical, and geological sciences. Despite the vast existing knowledge about the seismic waves produced by landslides on failures, much remains to be discovered about the dynamics responsible for these events. One way to improve our understanding of seismic activity, which have their starting point in inner layers of our planet, is the analysis of temporal series of events, from which, the probability of earthquakes occurrences may be calculated \cite{gutenberg1942,omori1894}.

Several studies have examined the phenomena under the viewpoint of complex systems, where, from nonlinear interactions between the elements of a system, complex patterns arise. In that direction, previous investigations, utilizing real data from seismic catalogs, and synthetic data from earthquake models, have analyzed spatio-temporal properties of seismicity from the perspective of non-extensive statistical mechanics \cite{abe2003law,darooneh2010,chochlaki2018global} and also using complex network theories \cite{peixoto2004distribution,tanaka2017statistical,pasten2018non}, which also have been applied to better understand many biological \cite{barabasi2004}, social \cite{newman2001}  and technological systems \cite{pastor2001}. It is important to note that non-extensive statistical mechanics, introduced in 1988 by Tsallis \cite{tsallis1988possible}, have been strongly used to argue in favor of long-range correlations in seismic events, as can be shown by the efforts made in \cite{sarlis2010nonextensivity,abe2005complexity,varotsos2005some}.

%\revision{It is important mentioning that efforts made by some scientists have showed the presence of non-extensive 
%used cto generalized some fundamental properties of } 

Abe and Suzuki \cite{abe2004scalefree,abe2004small} studied the complexity of seismic events by introducing the concept of {\it earthquakes network} that they built using networks of geographical sites by taking data of successive epicenters from seismological catalogs of some active regions. Although Abe and Suzuki named this network as \textit{earthquakes network}, in this paper we will adopt the nomenclature \textit{epicenters network}. We adopt a different name for two reasons: first because we believe that the latter name has a better agreement with the real sense of building the network; second to avoid confusion with another network that we will define later in this paper where nodes are actual quake events. The epicenters network in \cite{abe2004scalefree,abe2004small} was constructed by choosing a certain region of the world (e.g., California, Japan) and its respective earthquakes catalog, which gives for each seismic event, the magnitude, and a set of spatial and temporal data of the hypocenter. The geographic region they consider is then divided into small cubic cells, where a cell will become a vertex of the network if an earthquake has its epicenter therein, and two cells will be connected by a directed edge if two successive events occur in those respective cells. If two successive events occur in the same cell we have a self-edge. The network resulting from this process has been found to have non-trivial characteristics, being scale-free and small-world \cite{abe2006complex}. It is noteworthy that the same construction was made in \cite{peixoto2004distribution,peixoto2004statistics,peixoto2006network} but by using the known model for earthquakes dynamics proposed by Olami-Feder-Christensen (OFC model), which use concepts of self-organized criticality in non-conservative systems \cite{olami1992,christensen1992scaling}.

We argue however that the important issue regarding the possibility of long-range relationships between events located spatially and temporally far apart remains under-studied and almost unexplored. This paper provides evidences for such relationships.

\section{Method}
\label{method}

Observations reported from the analysis of the great 1906 San Francisco earthquake, in California, suggest that this earthquake has induced earthquakes several hundred miles away from the rupture zone (zone of breakage)  \cite{steeples1996aftershocks}. Thus, the possibility of seismic spatial long-range correlations emerges resulting in the analysis restricted to small seismic areas being inappropriate or incomplete. Moreover, studies from a temporal series of earthquakes, using different ways of building the network of epicenters suggest the existence of long-range correlations between different temporal and spatial seismic events, making it inconsistent with the hypothesis of the so called aftershocks zones \cite{baiesi2004scalefree,baiesi2005complex,abe2012universal,bendick2017weak}.

\begin{figure}[tb]
\begin{center}
\includegraphics[width=1.0\columnwidth]{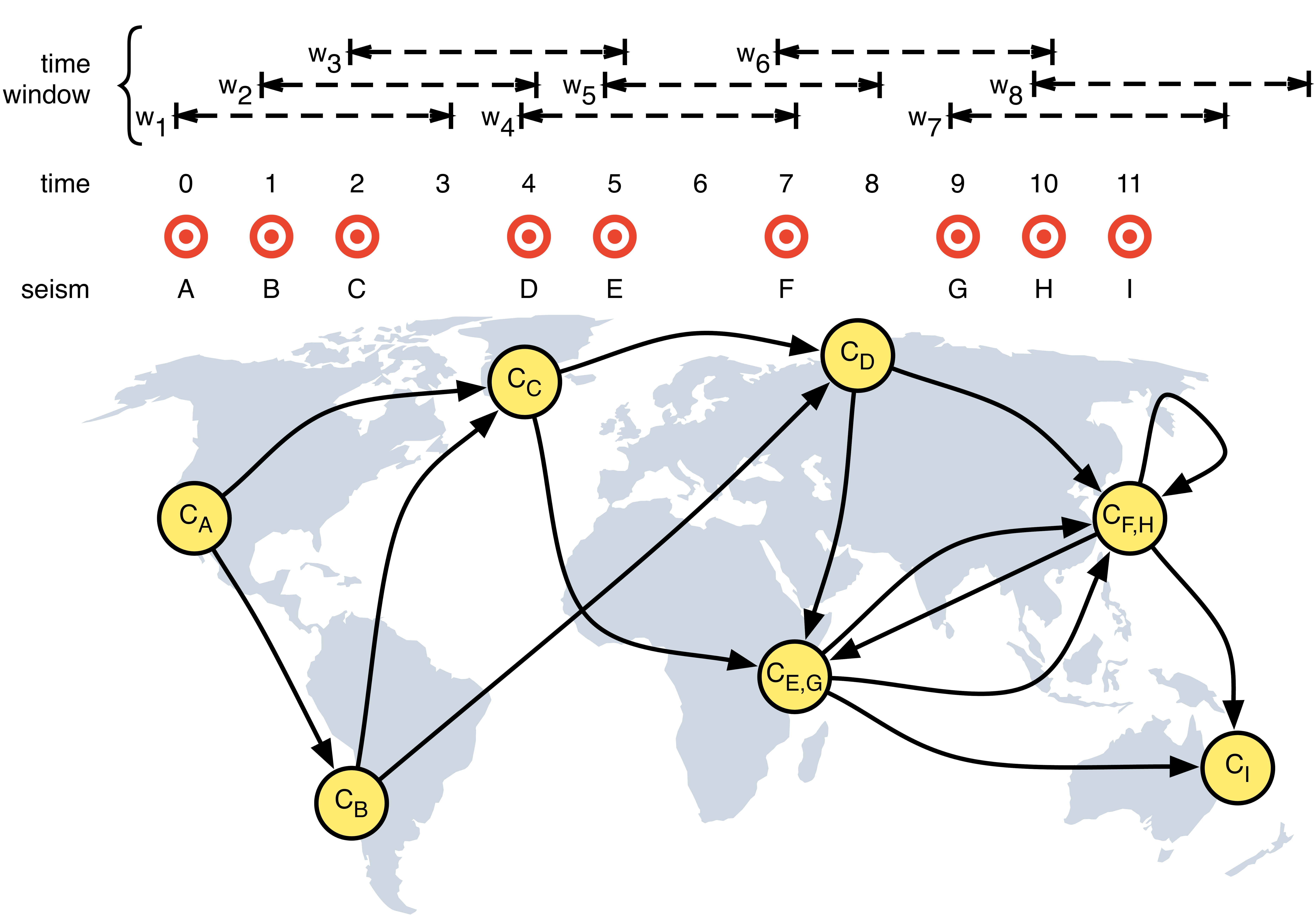}
\caption{Construction of the epicenters network. The time windows are represented by $w_i$, where $i$ is the window number and all time windows must have the same value, in this example $T$\,=\,3 (the time is represented in arbitrary units). Events in the same window are connected as explained in the text. We can see that there are 9 earthquakes ($A$, $B$, $C$, $D$, $E$, $F$, $G$, $H$, $I$), but the network of epicenters has only 7 vertices ($c_{A}$, $c_{B}$, $c_{C}$, $c_{D}$, $c_{E}$, $c_{F}$, $c_{I}$), because $c_{E} = c_{G}$ and $c_{F} = c_{H}$. It can also be observed that $c_{F} \rightarrow c_{H}$ is a self edge.}
\label{fig:example_net}
\end{center}
\end{figure}

%% I changed the greek letters to use the standards used for latitute (Phi) and 
%% Longitude (Lambda).

Using the same mechanism of connections for successive epicenters employed in \cite{abe2004scalefree,abe2006complex} for the construction of the epicenters network, but with data from the global catalog of earthquakes, Ferreira \et~\cite{ferreira2014small} built a global network of epicenters, considering all earthquakes with magnitude greater than 4.5 between the years 1972 to 2011, regardless of where in the world these earthquakes have taken place. The fundamental finding was that this global network is also complex, scale-free and small-world, which can lead to the interpretation that the growth rule of this network is of the preferential attachment type as described in \cite{barabasi1999emergence}. However, when analyzing the results obtained for the distribution of connectivity for this global network, created from the model of successive events, it was observed that the power-law behavior is maintained only for vertices with low connectivity presenting an exponential cutoff for the vertices of higher connectivity \cite{ferreira2014small}. This feature indicates a loss in the connectivity (preference) of vertices with a large number of connections. Therefore, in order to corroborate the studies on seismic dynamics through the use of theories of complex networks, in this paper we present a new method for building the global network of epicenters.

First, to make possible the definition of vertices in the network to be constructed, the surface of the planet is divided in equal square cells of side $L \times L$, where a cell will become a vertex of the network every time the epicenter of an earthquake (with depth smaller than 70\,km)  is located therein. Differently from the work of Abe and Suzuki \cite{abe2004scalefree,abe2006complex} and similar to our previous work \cite{ferreira2014small}, our cells are always $L \times L$ regardless of the position in the globe (done using Equation \ref{S}). To create the links between the vertices we use a chronologically ordered time series data of earthquakes and then define a ``time window'' ($w$) where the vertex corresponding to the first event is connected to all vertices within this window by directed edges but respecting the time order of earthquakes. Thereafter, the time window is moved forward so that we restart again at the next event and the new first vertex will be connected to all vertices within this window. To illustrate, suppose that it has been adopted a time window of value equal to $T$ in a given dataset of seismic events and that within the first window there are $N_s$ earthquakes ($s_1, s_2, \ldots ,s_{N_s}$), where each event occurs in a cell in the globe (not necessarily distinct). Thus, we assume that there is a probable relationship between the earthquake occurred in cell $c_{s_1}$ and the others $N_{s} - 1$ earthquakes within this time window, where this relationship will be represented in the network by directed edges between $c_{s_1} \rightarrow c_{s_2}$, $c_{s_1} \rightarrow c_{s_3}$, $\ldots$, $c_{s_1} \rightarrow c_{s_{N_{s-1}}}$, $c_{s_1} \rightarrow c_{s_{N_s}}$. Note, however, that due to the temporal sequence we do not have a direct edge added from $c_{s_k}$ to $c_{s_j}$ when $k > j$; i.e. an earthquake $s_k$ that happens after $s_j$ will produce the edge $c_{s_j} \rightarrow c_{s_k}$ but not the edge $c_{s_k} \rightarrow c_{s_j}$. After all the edges are added within one time window, the window will be moved forward, starting with the cell $c_{s_2}$, where the earthquake $s_2$ occurred. Then the same procedure for adding edges described above is repeated. The construction of the network will be completed when it is no longer possible to move the window forward. 
It means that, by introducing a ``time window" and connecting all the sites within this window, we minimize the loss of connections produced by the use of the successive model and thereby improve the construction method of epicenters networks, allowing us to find results more faithful to the dynamics of seismic phenomena. Thus, the time window is an agent that acts in a way to improve the method of connections between the elements of the network, since the value of the window will define what elements will be connected.

It is worth noting that, as done in previous works \cite{abe2006complex,davidsen2005analysis}, we have also neglected any classification of foreshocks, main shocks and aftershocks, putting all the events on an equal footing. Fig.~\ref{fig:example_net} illustrates an example of the process for creating this network of epicenters. It is also noted that if two or more epicenters belonging to the same time window occur in the same cell, this cell will be connected to itself forming a self-edge. Henceforth, we will use $k$ to refer to the in-degree of a node (its connectivity) in the network, and it represents the sum of the edge weights incoming a particular node.

The determination of the cell of each epicenter is made by observing the latitude $\phi_E$ and the longitude $\lambda_E$ of each epicenter to a given reference, which, for simplicity, has been adopted as $\phi_0 = 0$ e $\lambda_0 = 0$. From these coordinates we can calculate the north-south ($S_E^{ns}$) and east-west ($S_E^{ew}$) distances between the location of the epicenter and the referential adopted. Considering the spherical approximation of the Earth we have:
\begin{equation}
\label{S}
\begin{array}{l}   % they are two tiny letters "L"  inside the keys, to indicate "left-justified"
S^{ns}_E = R.\phi_E
\\
S^{ew}_E = R.\lambda_E.\cos{\phi_E},
\end{array}
\end{equation}

where $R = 6.371\times 10^3$ km is the Earth's radius. 

After calculating these distances, we need to define the size $L$ of the side of each square cell. It should be noted here that care must be taken when choosing the size of these cells, because if they are too small we will have a high resolution and very few earthquakes are likely to be in the same cell leading the network to be sparse and not quite useful. On the other hand, if the cells are too large the resolution will be too small, so there will be many repetitions of events within the same cell and the network will be too dense and also not useful. Following a previous study \cite{ferreira2014small}, in the present paper we will use $L \text{ = 20\,km}$, for the sides of the square cells $L \times L$ in the globe. Thus, to obtain the location of each cell in north-south and east-west directions, we have just to divide $S_E^{ns}$ e $S_E^{ew}$, respectively, by $L$.

\section{Data}
\label{data}

%%%%%%%%%%%%%%%%%%%%%%%%%%%%%%%%%%
\begin{figure}[tb]
\begin{center}
\subfigure[] {
\includegraphics[width=0.9\columnwidth]{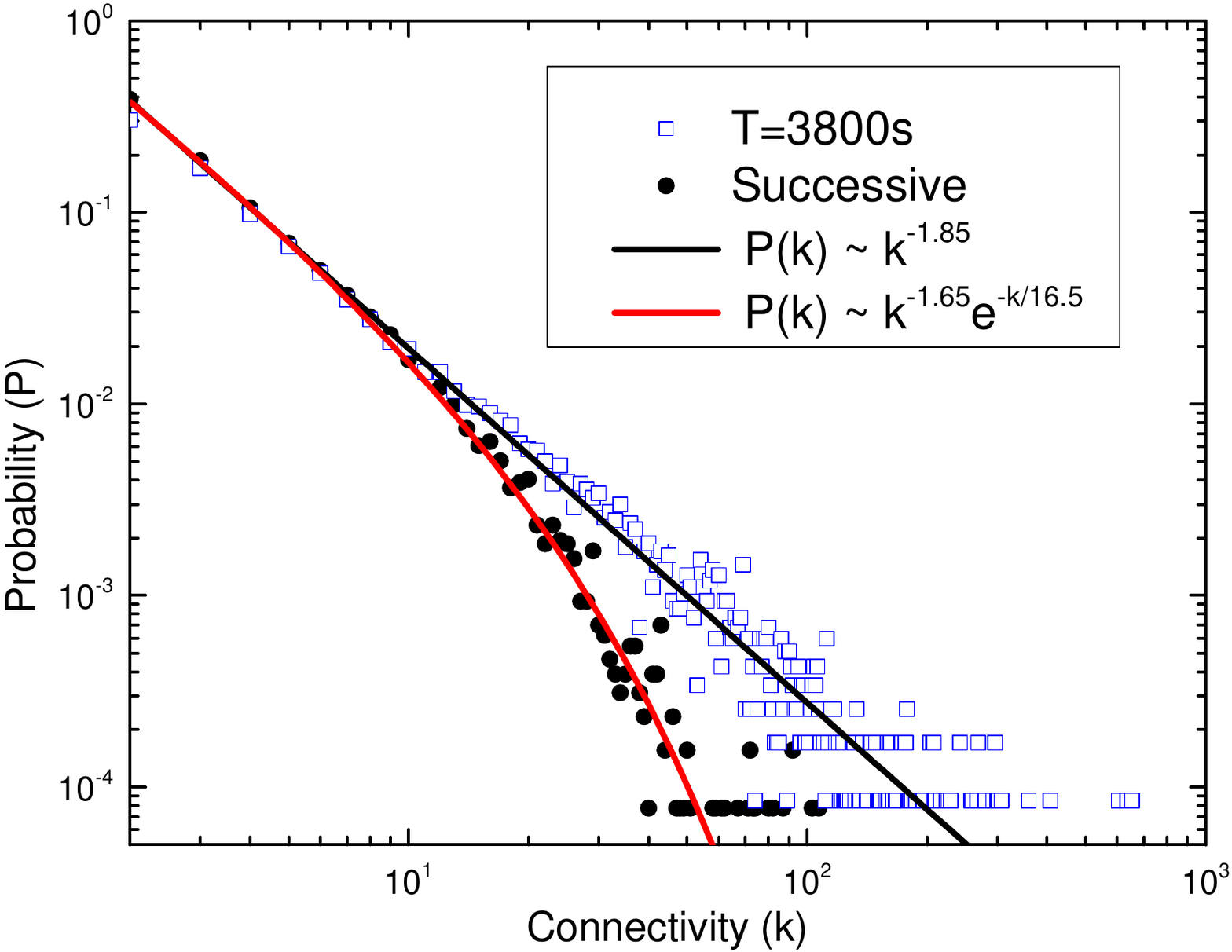}
\label{fig_P_win_nowin}
}
\subfigure[] {
\includegraphics[width=0.9\columnwidth]{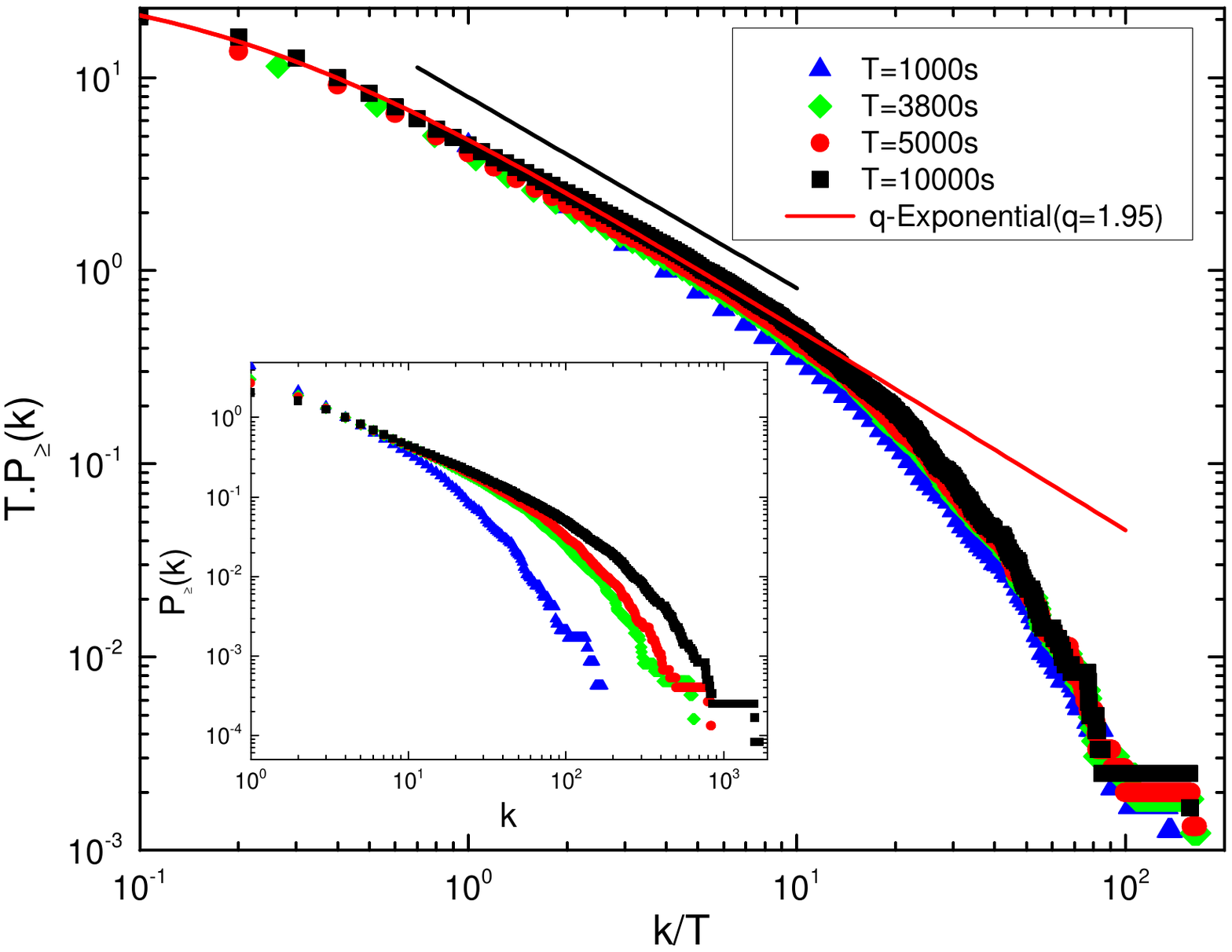}
\label{fig_P_win_scaling}
}
\caption{Probability connectivity distributions for World seismic data between 2002 and 2016 with $L$\,=\,20\,km. \subref{fig_P_win_nowin} Comparison between the time window model, for $T$\,=\,3\,800\,s, and  the model of successive events for the global network with $L$\,=\,20\,km. The model of successive events obeys a power-law with exponential cutoff as $P(k) \sim k^{-\delta}e^{-k/k_{c}}$, with $\delta=1.65$ and $k_{c}=16.5$, while the time window model one follows a power law $P(k) \sim k^{-\gamma}$, with $\gamma=1.85 \pm 0.01$. \subref{fig_P_win_scaling} Cumulative probability distributions, where by applying the scale function (\ref{scale_function}), we have a data collapse for $T$\,=\,1\,000\,s, 3\,800\,s, 5\,000\,s and 10\,000\,s. The best-fitting is for a $q$-exponential with $\beta=5.59 \pm 0.09$ and $q=1.95 \pm 0.01$ (red line). The black solid line with exponent $\alpha = 1.00$ is shown as a guide. \textit{Inset}: cumulative probability distributions without data collapse for $T$\,=\,1\,000\,s, 3\,800\,s, 5\,000\,s and 10\,000\,s.}
\label{fig:prob_nowin_win}
\end{center}
\end{figure}
%%%%%%%%%%%%%%%%%%%%%%%%%%%%%%%%%%

%%%%%%%%%%%%%%%%%%%%%%%%%%%%%%%%%%
\begin{figure*}[tb]
\begin{center}
\subfigure[] {
\includegraphics[width=0.615\columnwidth]{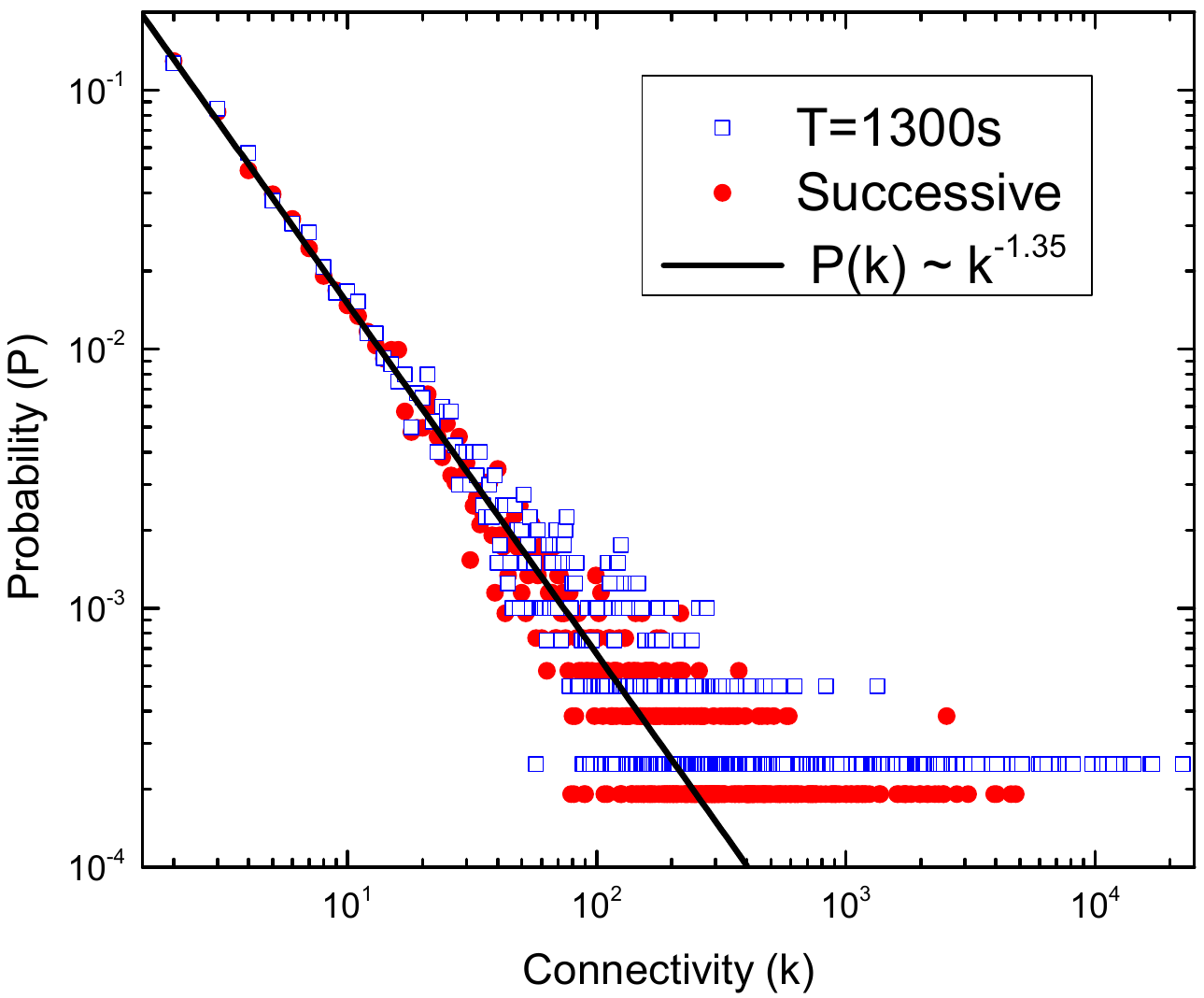}
\label{fig_P_win_nowin_SC}
}
\subfigure[] {
\includegraphics[width=0.66\columnwidth]{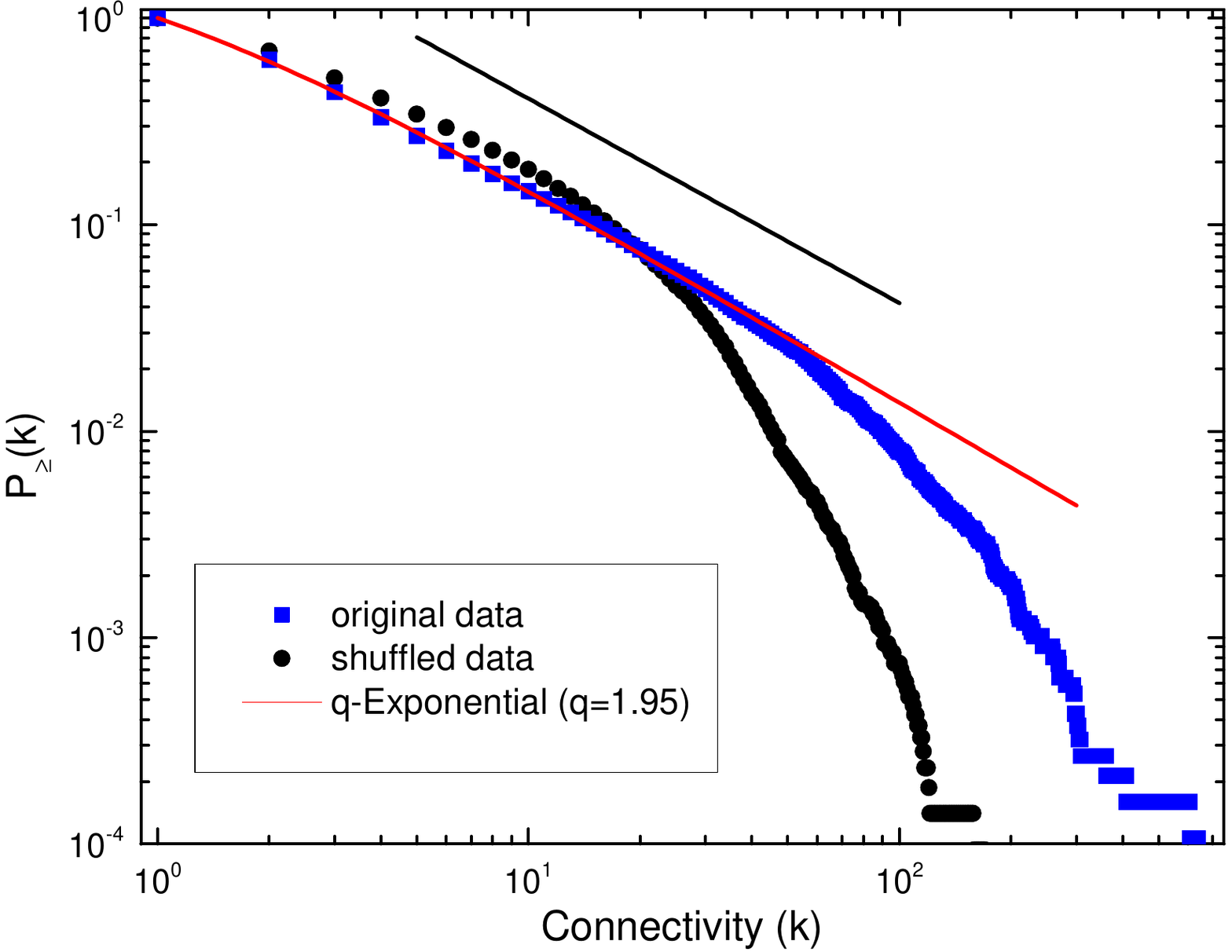}
\label{fig_P_shuffled}
}
\subfigure[] {
\includegraphics[width=0.66\columnwidth]{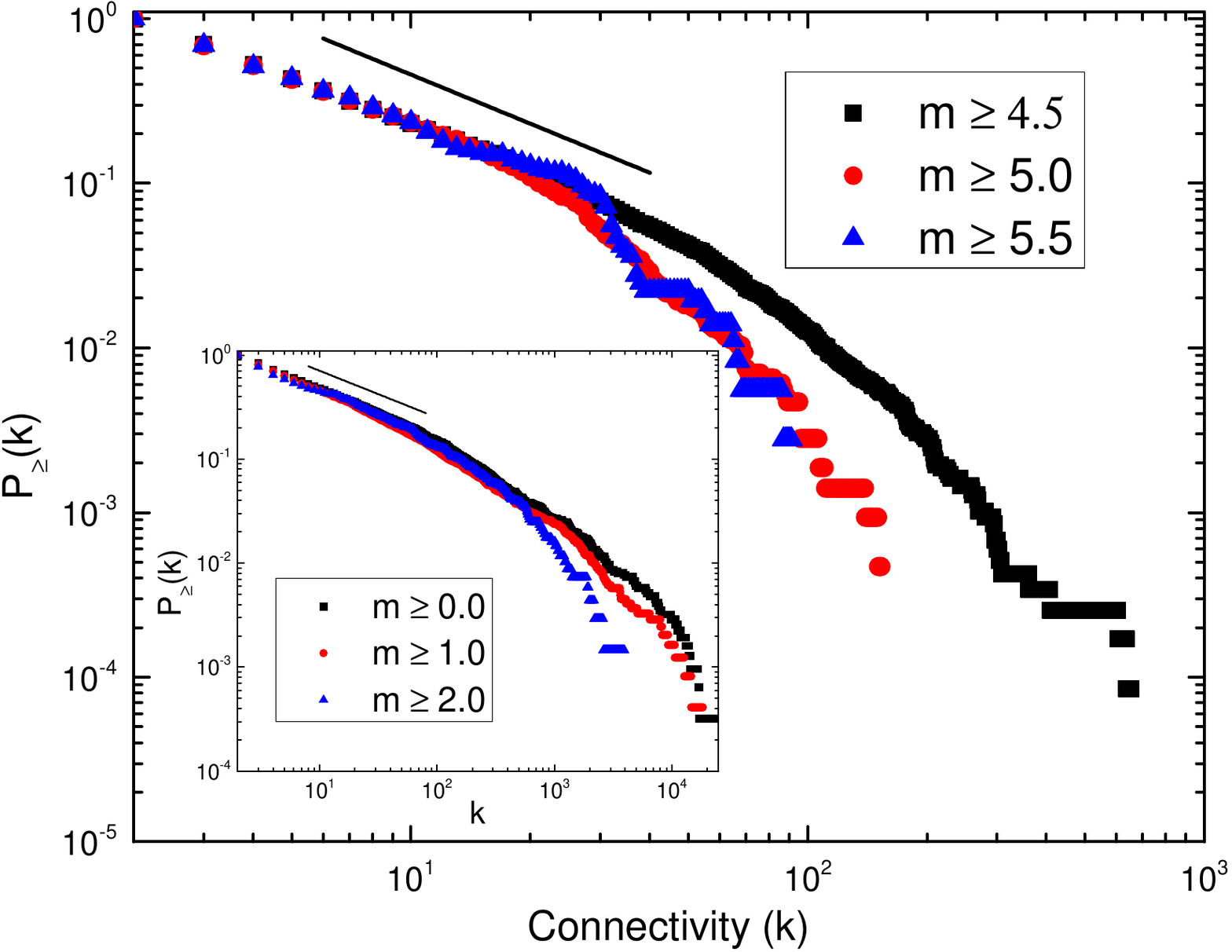}
\label{fig_P_mag}
}
\caption{Probability connectivity distributions for seismic data between 2002 and 2016. \subref{fig_P_win_nowin_SC} Comparison between the time window model, for $T$\,=\,1\,300\,s, and the model of successive events using the data for California with $L$\,=\,5\,km. Both models obey a power law with $\gamma=1.35 \pm 0.05$. \subref{fig_P_shuffled} Comparison between cumulative probability distributions of connectivities for global networks. The square (blue) represent the data for the network built using time-window model and the circle (black) represent the data for a shuffled network, both with $L$\,=\,20\,km and $T$\,=\,3\,800\,s. It is possible to see that the data obtained from the shuffled network do not follows a power law or a $q$-exponential functions. The black solid line has exponent $\alpha = 1.00$ and is shown as a guide. \subref{fig_P_mag} Cumulative probability distribution for World earthquakes with $m \geq 4.5$, $m \geq 5.0$ and $m \geq 5.5$. \textit{Inset}: cumulative probability distribution for California earthquakes with $m \geq 0.0$, $m \geq 1.0$ and $m \geq 2.0$. The black solid line has exponent $\alpha = 1.00$.}
\label{fig:tests}
\end{center}
\end{figure*}

%(e) Comparison between original data and data for a shuffled global network, both with $L$\,=\,20\,km and $T$\,=\,3\,000\,s. The red solid line has exponent $\alpha = 1.00$. (f) Cumulative probability distribution for earthquakes with $m \geq 4.5$, $m \geq 5.0$ and $m \geq 5.5$. The black solid line has exponent $\alpha = 1.00$. In all cases $k$ represents in-degree.
%%%%%%%%%%%%%%%%%%%%%%%%%%%%%%%%%%

The dataset used in our global analysis was obtained from the World Catalogue of Earthquakes of Advanced National Seismic System (ANSS) \footnote{http://quake.geo.berkeley.edu/anss} and includes earthquakes between 2002 and 2016. In this catalog, events whose magnitudes ($m$) are less than 4.5 (Richter scale) are not recorded for all parts of the world, so to obtain a more homogeneous distribution of the data for the whole world we consider only earthquakes with $m \geq 4.5$. In addition, in order to make comparisons only between events that have similar seismic origins, we will consider in this work only shallow earthquakes (i.e, earthquakes that have depth up to 70\,km), which allows us to use our model of approximation of square cells, rather than cubic. The total of events is 80\,520. The probability distribution of magnitudes of earthquakes in our data is in agreement with the Gutenberg-Richter law with a {\it{b-value}} exponent equal to $1.080 \pm 0.003$, which was obtained by the maximum likelihood method. This result is expected, given that the Gutenberg-Richter law only presents problems for small magnitude values \cite{darooneh2010}.

\section{Results}
\label{results}

After the construction of the network, we performed some experiments in order to understand the structure of this network. As stated earlier, studies using the model of successive worldwide connections, show a distribution of connectivities in the form of power laws with an exponential cutoff. However, when analyzing the same data stream using the standard window of time for various values of the window, we observe two important features in the distribution of connectivities: 
$(i)$~in Fig~\ref{fig_P_win_nowin} we can note that the exponential cutoff disappears in the density probability, recovering for the entire range of connectivity, the behavior type $P(k) \sim k^{-\gamma}$, leading again to the idea that the growth rule of this network follows a preferential attachment. In addition, we emphasize the fact that we also have calculated the $\gamma$ exponent using $L \text{ = 10\,km, 30\,km and 40\,km}$ and in all cases we have the same behavior.
$(ii)$~using the cumulative probability distribution we observe that the best fit is obtained by a non-traditional function, namely a $q$-exponential, $P_{\geq}(k) = A[1 - (1 - q)\beta k]^{1/(1 - q)}$, which is a generalization of the exponential curve, where taking the limit $q \rightarrow 1$ we recover the standard exponential and for $q > 1$ it exhibits power law asymptotic behavior (as we can see in Fig~\ref{fig_P_win_scaling} for larger values of $k$). However, it is noteworthy that it is possible to make the distribution invariant with respect to the value of the time window by using the scale function:
\begin{equation}
\label{scale_function}
P_{\geq}(k,T) = T^{-1}f(k/T), 
\end{equation}
where $f(x)$ decays as $x^{-\alpha}$ with $\alpha = 1.00$ and the data collapse is in agreement with the scaling hypothesis, as seen in Fig~\ref{fig_P_win_scaling}. 

An important point to highlight is that in non-extensive statistical mechanics the \textit{q}-exponential distribution appears naturally from the maximization of the Tsallis entropy \cite{tsallis1988possible}, which is used to explain many complex systems with characteristics such as long-range interaction between its elements and long-range temporal memory, where the traditional Boltzmann-Gibbs statistical mechanics does not seem to apply \cite{koutalonis2017evidence,vallianatos2013evidence,barbosa2013statistical,ludescher2011universal,aad2011charged,tamarit1998sensitivity}. Another point of interest is that a previous work has found the non-extensive behavior in networks of epicenters using a modified version of the OFC model, where the small-world effect was taken into account in the lattice topology construction \cite{ferreira2015agreement}.

In order to show the consistency of the time window methodology, we performed three tests. 
The first is based on the fact that, as stated above, for small active regions on the planet, the model of successive events produce a connectivity distribution followed by a power-law function without cutoff \cite{abe2003law,abe2004scalefree}, so, for consistency, for small regions, the time window model must have similar results. 
To demonstrate this we have used data obtained from the catalog of Southern California Earthquake Data Center (SCEDC) \footnote{http://scedc.caltech.edu/eq-catalogs/}, for earthquakes with magnitudes $m \geq 0$ between 2002 and 2016, to plot the graph of the distribution of connectivities for the network of epicenters of California ($30^\circ \mathrm{N} - 38^\circ \mathrm{N}$ and $113^\circ \mathrm{W} - 122^\circ  \mathrm{W}$), where the total number of events was 231\,612 and the cell size considered was  $L$\,=\,5\,km, as previously used in \cite{abe2004scalefree,abe2006complex,ferreira2014small}. As can be seen in Fig.~\ref{fig_P_win_nowin_SC}, for California both models have similar results, as it should be. 
The second test was to take the time series data and randomly rearrange the locations where the events occurred, while maintaining the time instant they occurred. This test has the objective to check out if our findings have occurred by chance or not, i.e, we are checking to see if our results change when we change randomly the locations of the epicenters. Figure~\ref{fig_P_shuffled} shows that the distribution of connectivities for the network constructed from the data randomly rearranged is very different from the original distribution using the time-window model, which means the sequential order of the event matters, i.e, since the shuffled network has not the same characteristics found in the network built using by applying the model, the phenomena is not governed by an aleatory behavior. 
The last test is similar to a completeness test, i.e, we check if the value that we considered as lower threshold of magnitude ($m=4.5$) is satisfactory and if it does not interfere in the results. To do that we have analyzed the connectivity distribution using different magnitude thresholds for the global seismic data. The magnitude intervals considered were $m \geq 4.5$, $m \geq 5.0$ and $m \geq 5.5$, comprehending 80\,520, 22\,732 and 6\,297 earthquakes respectively, with the number of nodes in the epicenters network in each case equal to 23\,380, 6\,184 and 1\,243. We have also performed the same test using the California data, considering $m \geq 0.0$, $m \geq 1.0$ and $m \geq 2.0$, with 231\,612, 143\,012 and 30\,303 earthquakes and the number of nodes equal to 4\,591, 4\,004 and 1\,377, respectively.
As we can see in Fig.~\ref{fig_P_mag}, in all cases we have the same power law behavior, but the higher the number of nodes in the network the larger is the power law regime, as expected. 

%%%%%%%%%%%%%%%%%%%%%%%%%%%%%%%%%%
\begin{figure}[t]
\begin{center}
\subfigure[] {
\includegraphics[width=0.65\columnwidth]{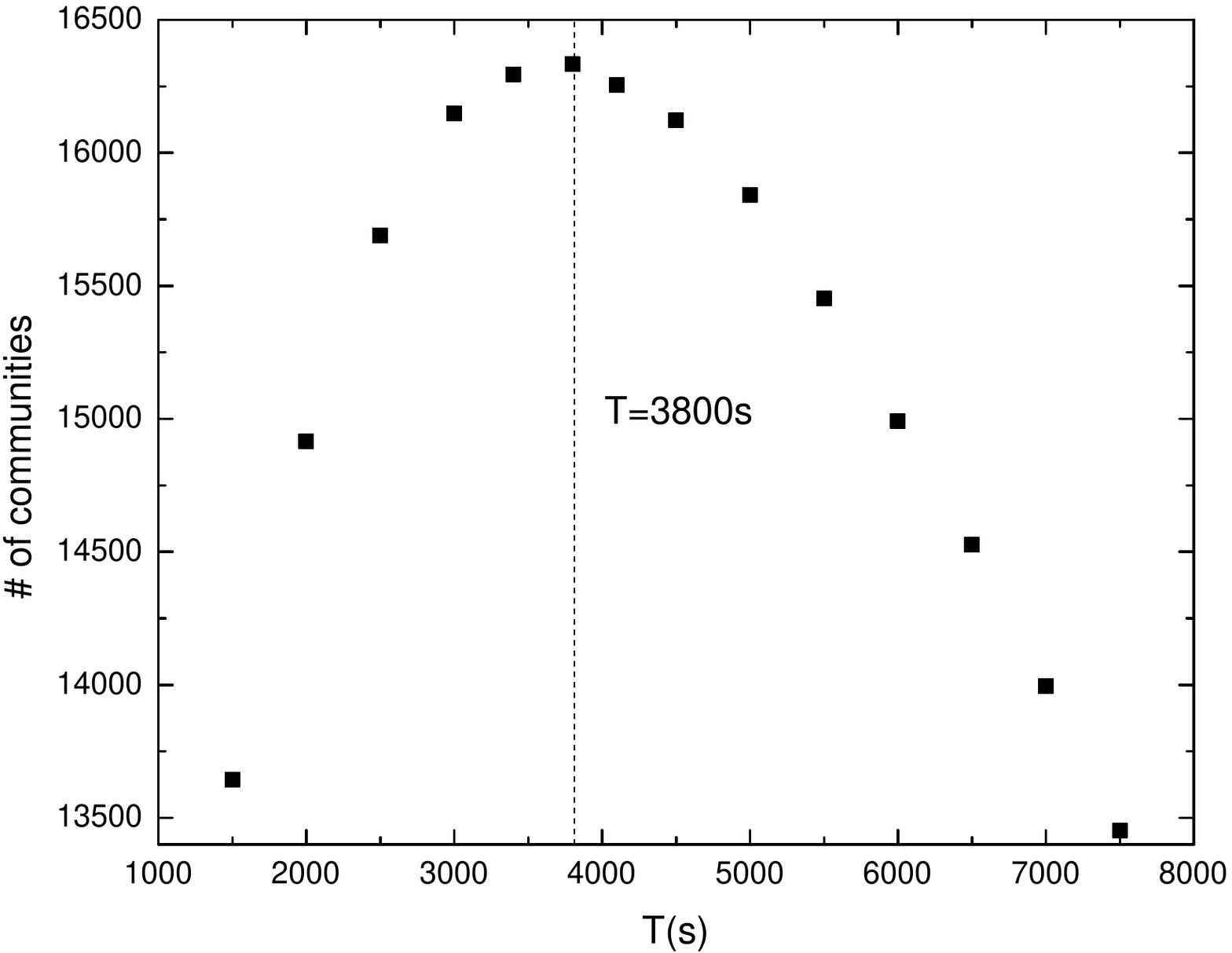}
\label{win_communities_world}}
\subfigure[] {
\includegraphics[width=0.65\columnwidth]{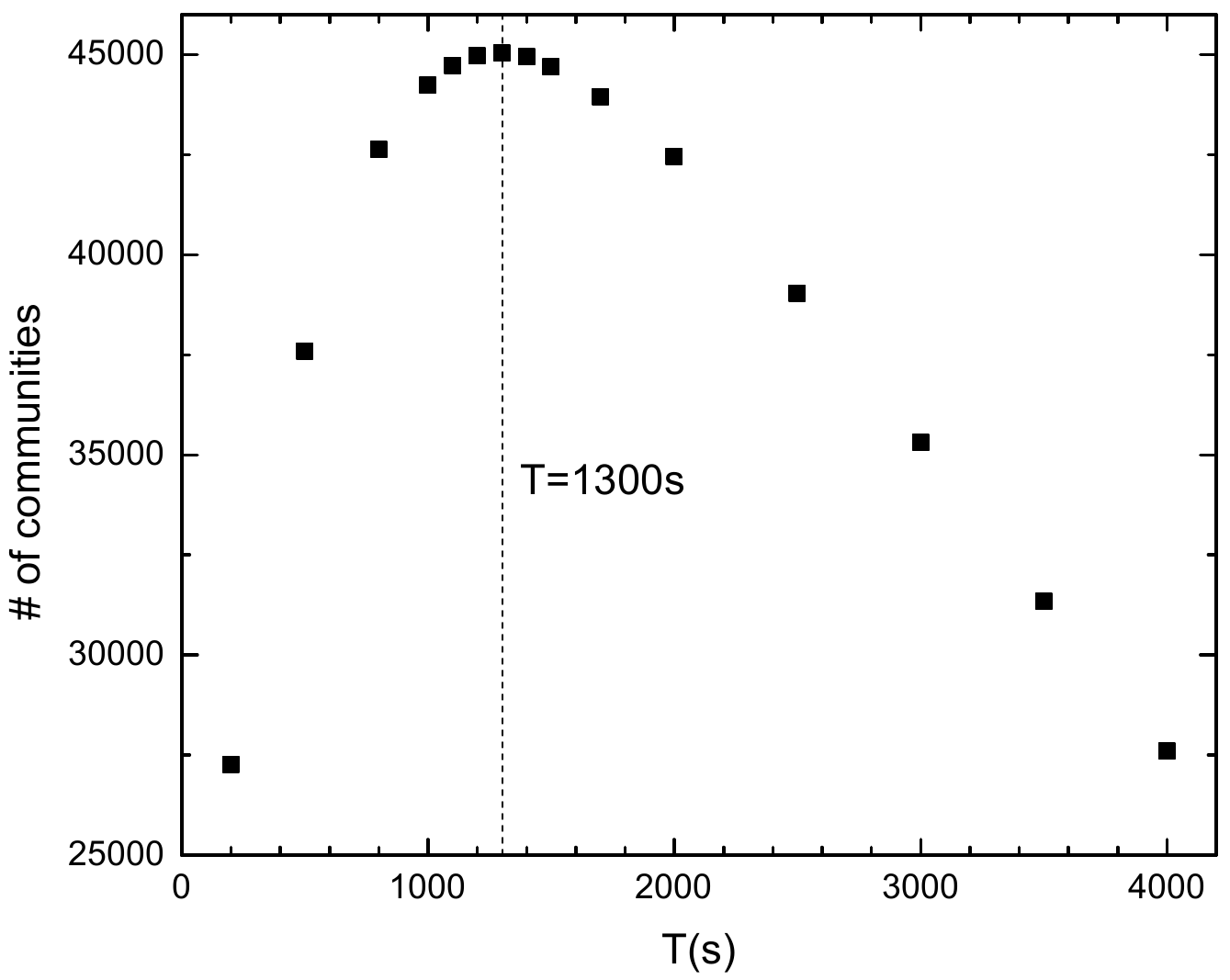}
\label{win_communities_california}}
\caption{Variation of the number of communities in the earthquakes network with the size of the time window. Seismic data between the years 2002 and 2016. The best window is found when the number of communities is maximum, i.e., \subref{win_communities_world} at $T$\,=\,3\,800\,s, with $m \geq 4.5$, for the world case, and \subref{win_communities_california} at $T$\,=\,1\,300\,s, with $m \geq 0$  for the California case. }
\label{fig:communities}
\end{center}
\end{figure}
%%%%%%%%%%%%%%%%%%%%%%%%%%%%%%%%%%

%%%%%%%%%%%%%%%%%%%%%%%%%%%%%%%%%%
\begin{figure*}[t]
\begin{center}
\subfigure[] {
\fbox{\includegraphics[width=0.94\columnwidth]{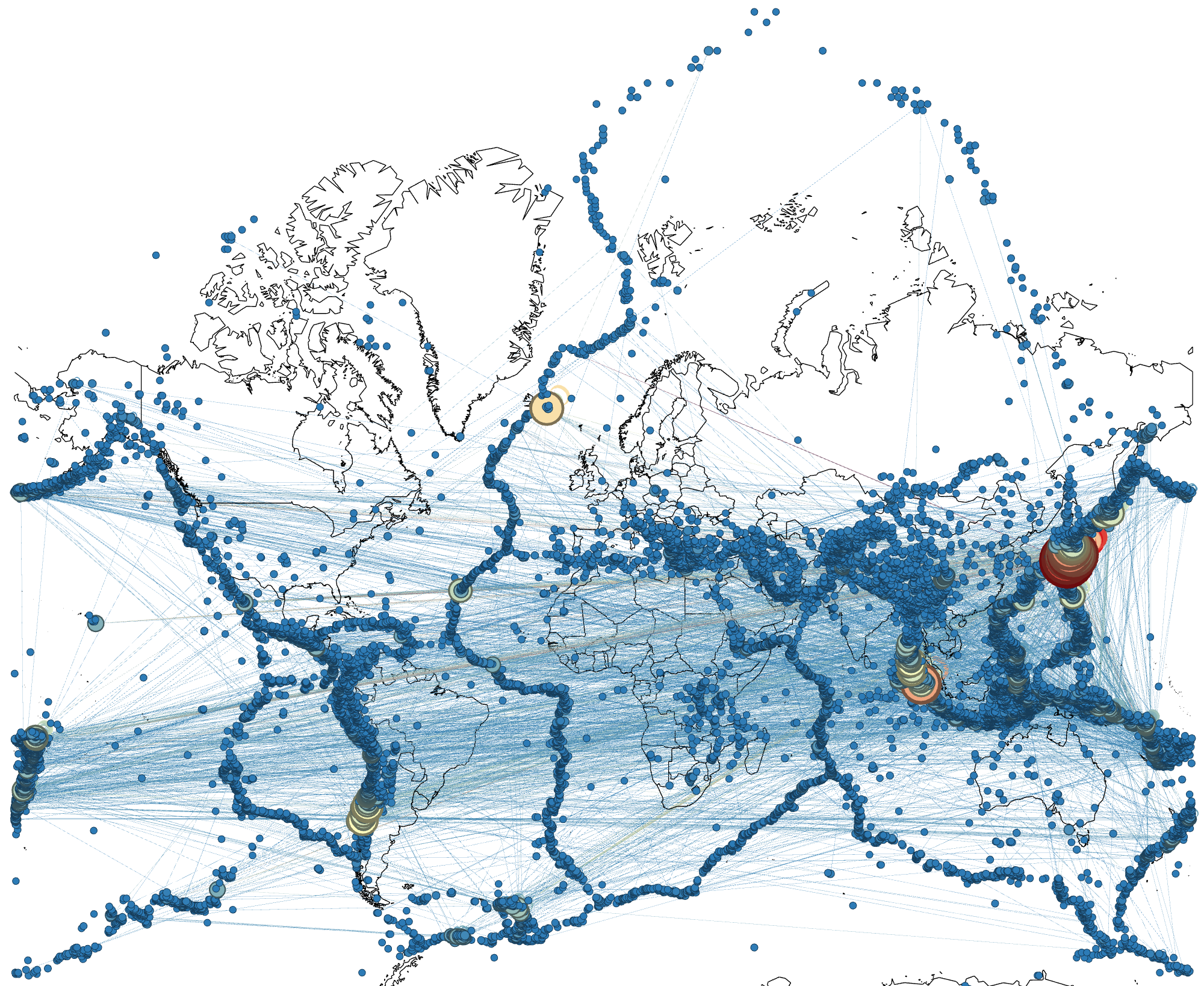}}
%\fbox{\includegraphics[width=0.74\columnwidth]{pics/fig3b_mapoverlay.pdf}}
\label{gephi_timewindow}}
\subfigure[] {
\fbox{\includegraphics[width=0.94\columnwidth]{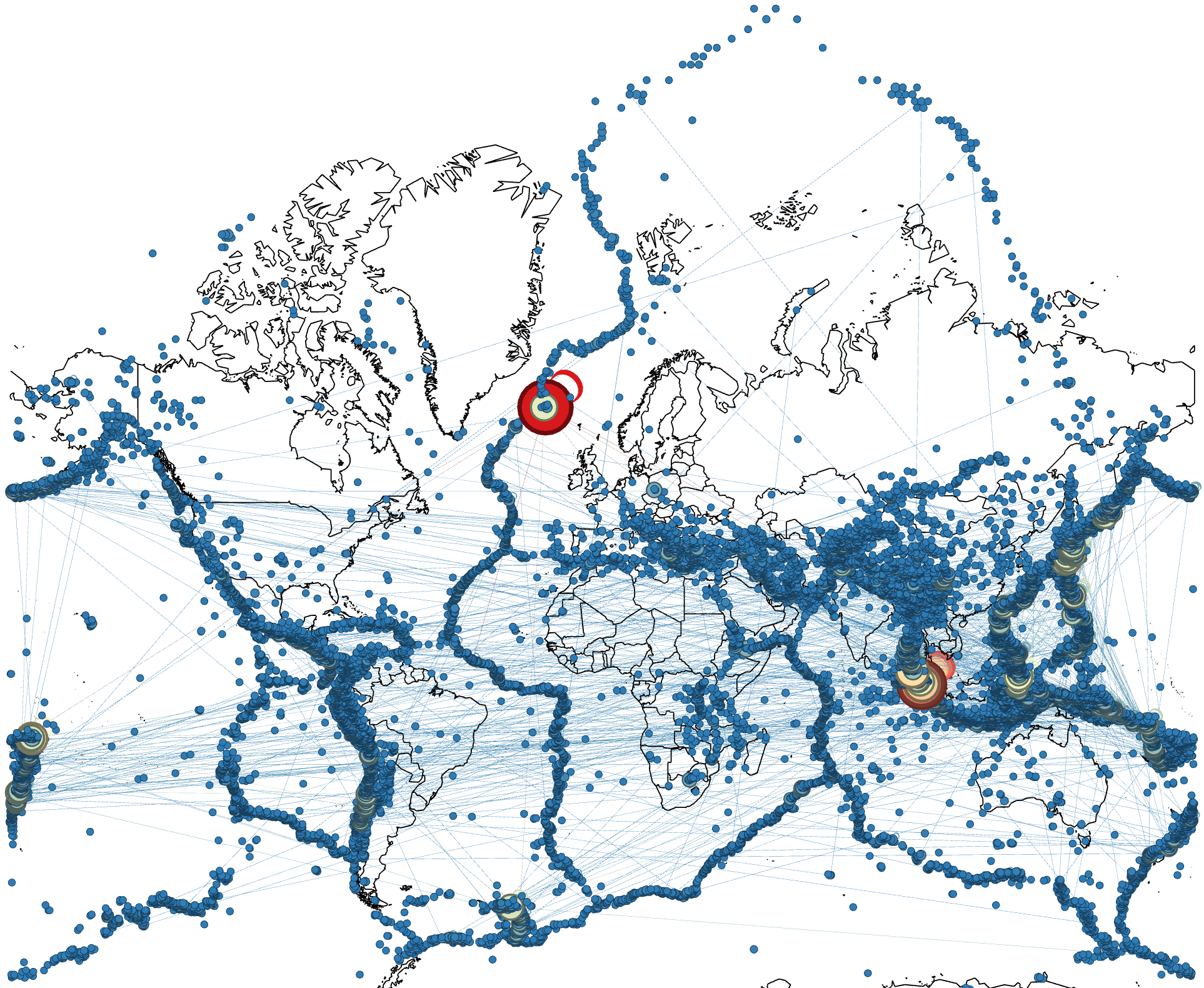}}
\label{gephi_successive}}
\caption{Geospatial picture of the global network of cells between the years 2002 and 2016, for $L$\,=\,20\,km and $m \geq 4.5$, where it is possible to clearly observe the tectonic faults delineate by cells in the network. For clarity, only the links that occurred more than 2 times between the same two cells are shown. Larger and reddish cells have higher number of connections. \subref{gephi_timewindow}~Time-window model for $T$\,=\,3\,800\,s. The sites with the largest cells are located around Japan, Sumatra Island, Chile and Iceland, in that order. \subref{gephi_successive}~Successive model. The largest cells are located in a different order.}
\label{fig:gephi_net}
\end{center}
\end{figure*}
%%%%%%%%%%%%%%%%%%%%%%%%%%%%%%%%%%

To better characterize our network of epicenters it is important to study, in addition to the distribution of connectivities, other features of this network. Two important metrics in the study of complex networks are the {\it{clustering coefficient}} ($C$) and the {\it{average shortest path}} ($\ell$). Their importances stem from the fact that they can be used to characterize small-world networks. Small-world networks are those with dense connectivity areas and long jumps between these areas. Hence, to be considered small-world a network needs to have a small {\it{average shortest path}} when compared to the number of vertices and a high {\it{clustering coefficient}} compared to a similar random network. To perform the analysis of the {\it{clustering coefficient}}  and the {\it{average shortest path}} in our network, one must recall that the chosen time window will directly affect the measure of these quantities. Thus, it is natural to wonder if there is a best value for the time window. Prior to answering this question, remember that if we are trying to insert a time window in the time series data, this window will be inserted in the \textit{earthquakes network}, i.e., in the network of connections between earthquakes (in the example given earlier the vertices of this network are: $s_1, s_2, \ldots ,s_{N_s}$) and not in the \textit{epicenters network}, which is the spatial network of cells where earthquakes occurred (in the example previously given the vertices of this network are: $c_{s_1}, c_{s_2}, \ldots ,c_{s_{N_s}}$). We also emphasize the impossibility of loops in the earthquakes network, since an event cannot be connected to itself (different from the epicenters network, where a cell can be linked to itself, as illustrated in Fig.\,\ref{fig:example_net}). So we need to find a window not too large, that will create many links between earthquakes that probably are not correlated, but neither too small, which would bring to us a very fragmented network in which many earthquakes that have correlations between them are not connected. Thus, to find this window, we have calculated the number of communities in the networks of earthquakes and observed how this number depends on window size. In simple terms, a community is a collection of vertices in the network that is densely connected internally (inside the community). To calculate the number of communities we use the Louvain method \cite{blondel2008fast}. From the results shown in Fig.\,\ref{fig:communities}, we can confirm the fact that small windows will cause the network of earthquakes to be very fragmented and consequently with few communities, and large windows will produce very large clusters also causing a decrease in the number of communities. 
This leads us to the conclusion that the ideal window occurs when the number of communities is a maximum and, for the global earthquakes network, this happens, when $T$\,=\,3\,800\,s [Fig.\,\ref{win_communities_world}]. At this point it is important highlight that the time window value found do not break any physical or geophysical property, since a seismic wave propagating on the surface of the Earth during that time interval will only cover a small fraction of the Earth's surface. Applying the same procedure for the California earthquakes network, we find the ideal window for $T$\,=\,1\,300\,s [Fig.\,\ref{win_communities_california}].

Using the ideal time window found, we built a global network of epicenters (spatial network of cells where the epicenters have occurred) using cells of size 20\,km $\times$ 20\,km. The measurements of $C$ and $\ell$, were performed using the algorithms described in \cite{barrat2004architecture} and \cite{brandes2001faster}, respectively. The results for our epicenters network (not to be confused with the earthquakes network definition of Abe and Suzuki \cite{abe2004scalefree,abe2004small}) were $C=5.18 \times 10^{-1}$ and $\ell=5.24$. For a random network with the same size the \textit{clustering coefficient} has a $C_{rand}=3.34 \times 10^{-4}$ value. From these results we observe that the global network of epicenters created from our model of time windows has small-world properties, given that $C \gg C_{rand} $ and that $\ell \ll N$ ($\ell \approx \ln N$).

In our network construction it is possible to know the exact geographical location of each cell, the number of connections of each cell and to which other cells they are connected to. Therefore we used the software Gephi \footnote{https://gephi.org/} to obtain a geo-spatial image of the earthquakes. In Fig.~\ref{gephi_timewindow} it is possible to observe that the cells of higher connectivity are located in Japan, in the Sumatra Island, in Chile and in Iceland. This fact is interesting because between 2002 and 2016 these regions were those with earthquakes of very elevated magnitudes or very intense seismic activities (as in the Iceland case, due to volcanic eruptions, especially between 2014 and 2015)\,\footnote{https://earthquake.usgs.gov/earthquakes/search/}, which makes sense since it is expected that earthquakes of great magnitudes produce more aftershocks than earthquakes of smaller magnitudes. 
Note however that this result is not found when using the model of successive events, where the Japan region for example, despite having been the region with larger earthquakes is not the region of largest connectivity, and has connectivity similar to that found in the New Zealand region [Fig.~\ref{gephi_successive}]; once again, we have a good indication that the time-window model is a better approach for constructing networks of epicenters.

\section{Conclusions}
\label{conclusions}

We showed that the global network of epicenters obtained from the time-window construction has scale-free properties, where the distribution of connectivities follows a \textit{q-exponential} and, if using a scaling function, it is shown to be invariant with respect to the value of time window adopted. We also showed the consistency of the construction by three methods: first, by applying the model to a small region (California) and finding similar results to those found when using the model of successive events; second rearranging the time series data of epicenters and finding different results from those found when using original data; third, we found the same behavior for the connectivity distribution by using different magnitude thresholds. Furthermore, we have defined a mechanism for determining the best time window and observe that the network built using that window has small-world characteristics and the cells with the greatest connectivity are located in Japan, Sumatra, Chile and Iceland (regions with very intense shallow seismic activities from 2002--2016). This shows that our method of network construction is able to naturally identify the regions of higher seismic intensities using only properties of complex networks and that, through the existing connections in the network, it becomes possible to establish relations between the most diverse regions of the world. Thus, due to the presence of small world properties and to the non-extensive characteristics found, our results constitute evidences of possible long-range correlations between spatially-separated locations as well as a long-range temporal memory between earthquakes temporally apart from each other.

Given the focus on shallow earthquakes in this paper, a natural future work is to verify the suitability of the same model for deep earthquakes and understand how their inclusion affect the structure of the epicenters network. Note that we have used square regions but the consideration of deep earthquakes may indeed require the use of cubic regions.

% \revision{OBS: trabalho para grandes profundidades, "deep-focus earhquakes", verificar se o modelo de célula quadrada é uma boa aproximação, ou precisam ser feitas céulas cúbicas.}

\acknowledgments
ARRP thanks the CNPq (Brazilian Funding Agency), for a research fellowship.

\bibliographystyle{eplbib}
\bibliography{bib}

\begin{thebibliography}{10}
\expandafter\ifx\csname url\endcsname\relax\def\url#1{\texttt{#1}}\fi

\bibitem{gutenberg1942}
\Name{Gutenberg B. \and Richter C.} \REVIEW{Bull. Seism. Soc.
  Am.}{32}{1942}{163}.

\bibitem{omori1894}
\Name{Omori F.} \REVIEW{Coll. Sci. Imper. Univ. Tokyo}{7}{1894}{111}.

\bibitem{abe2003law}
\Name{Abe S. \and Suzuki N.} \REVIEW{J. Geophys. Res.}{108}{2003}{2113}.

\bibitem{darooneh2010}
\Name{Darooneh A.~H. \and Mehri A.} \REVIEW{Physica A: Statistical Mechanics
  and its Applications}{389}{2010}{509}.

\bibitem{chochlaki2018global}
\Name{Chochlaki K., Vallianatos F. \and Michas G.} \REVIEW{Physica A:
  Statistical Mechanics and its Applications}{493}{2018}{276}.

\bibitem{peixoto2004distribution}
\Name{Peixoto T.~P. \and Prado C.~P.} \REVIEW{Physical Review
  E}{69}{2004}{025101}.

\bibitem{tanaka2017statistical}
\Name{Tanaka H. \and Hatano T.} \REVIEW{The European Physical Journal
  B}{90}{2017}{248}.

\bibitem{pasten2018non}
\Name{Past{\'e}n D., Torres F., Toledo B.~A., Mu{\~n}oz V., Rogan J. \and
  Valdivia J.~A.} \REVIEW{Physica A: Statistical Mechanics and its
  Applications}{491}{2018}{445}.

\bibitem{barabasi2004}
\Name{Barab{\'a}si A.-L. \and Oltvai Z.~N.} \REVIEW{Nature Reviews
  Genetics}{5}{2004}{101}.

\bibitem{newman2001}
\Name{Newman M.~E.} \REVIEW{Proceedings of the National Academy of
  Sciences}{98}{2001}{404}.

\bibitem{pastor2001}
\Name{Pastor-Satorras R., V{\'a}zquez A. \and Vespignani A.} \REVIEW{Physical
  Review Letters}{87}{2001}{258701}.

\bibitem{tsallis1988possible}
\Name{Tsallis C.} \REVIEW{Journal of Statistical Physics}{52}{1988}{479}.

\bibitem{sarlis2010nonextensivity}
\Name{Sarlis N., Skordas E. \and Varotsos P.} \REVIEW{Physical Review
  E}{82}{2010}{021110}.

\bibitem{abe2005complexity}
\Name{Abe S., Tirnakli U. \and Varotsos P.} \REVIEW{europhysics
  news}{36}{2005}{206}.

\bibitem{varotsos2005some}
\Name{Varotsos P., Sarlis N., Tanaka H. \and Skordas E.} \REVIEW{Physical
  Review E}{71}{2005}{032102}.

\bibitem{abe2004scalefree}
\Name{Abe S. \and Suzuki N.} \REVIEW{Europhys. Lett.}{65}{2004}{581}.

\bibitem{abe2004small}
\Name{Abe S. \and Suzuki N.} \REVIEW{Physica A}{337}{2004}{357}.

\bibitem{abe2006complex}
\Name{Abe S. \and Suzuki N.} \REVIEW{Nonlin. Processes
  Geophys.}{13}{2006}{145}.

\bibitem{peixoto2004statistics}
\Name{Peixoto T.~P. \and Prado C.~P.} \REVIEW{Physica A: Statistical Mechanics
  and its Applications}{342}{2004}{171}.

\bibitem{peixoto2006network}
\Name{Peixoto T.~P. \and Prado C.~P.} \REVIEW{Physical Review
  E}{74}{2006}{016126}.

\bibitem{olami1992}
\Name{Olami Z., Feder H. J.~S. \and Christensen K.} \REVIEW{Physical Review
  Letters}{68}{1992}{1244}.

\bibitem{christensen1992scaling}
\Name{Christensen K. \and Olami Z.} \REVIEW{Physical Review A}{46}{1992}{1829}.

\bibitem{steeples1996aftershocks}
\Name{Steeples D.~W. \and Steeples D.} \REVIEW{Bull. Seismol. Soc.
  Am.}{86}{1996}{921}.

\bibitem{baiesi2004scalefree}
\Name{Baiesi M. \and Paczuski M.} \REVIEW{Phys. Rev. E}{69}{2004}{}.

\bibitem{baiesi2005complex}
\Name{Baiesi M. \and Paczuski M.} \REVIEW{Nonlin. Processes
  Geophys.}{12}{2005}{1}.

\bibitem{abe2012universal}
\Name{Abe S. \and Suzuki N.} \REVIEW{Eurphys. Lett}{97}{2012}{}.

\bibitem{bendick2017weak}
\Name{Bendick R. \and Bilham R.} \REVIEW{Geophysical Research
  Letters}{44}{2017}{8320}.

\bibitem{ferreira2014small}
\Name{Ferreira D.~S., Papa A.~R. \and Menezes R.} \REVIEW{Physica A:
  Statistical Mechanics and its Applications}{408}{2014}{170}.

\bibitem{barabasi1999emergence}
\Name{Barab{\'a}si A.-L. \and Albert R.} \REVIEW{Science}{286}{1999}{509}.

\bibitem{davidsen2005analysis}
\Name{Davidsen J. \and Paczuski M.} \REVIEW{Phys. Rev. Lett.}{94}{2005}{}.

\bibitem{koutalonis2017evidence}
\Name{Koutalonis I. \and Vallianatos F.} \REVIEW{Pure and Applied
  Geophysics}{174}{2017}{4369}.

\bibitem{vallianatos2013evidence}
\Name{Vallianatos F. \and Sammonds P.} \REVIEW{Tectonophysics}{590}{2013}{52}.

\bibitem{barbosa2013statistical}
\Name{Barbosa C.~S., Ferreira D.~S., do~Esp{\'\i}rito~Santo M.~A. \and Papa
  A.~R.} \REVIEW{Physica A: Statistical Mechanics and its
  Applications}{392}{2013}{6554}.

\bibitem{ludescher2011universal}
\Name{Ludescher J., Tsallis C. \and Bunde A.} \REVIEW{EPL (Europhysics
  Letters)}{95}{2011}{68002}.

\bibitem{aad2011charged}
\Name{Aad G., Abbott B., Abdallah J., Abdelalim A., Abdesselam A., Abdinov O.,
  Abi B., Abolins M., Abramowicz H., Abreu H. \etal} \REVIEW{New Journal of
  Physics}{13}{2011}{053033}.

\bibitem{tamarit1998sensitivity}
\Name{Tamarit F., Cannas S. \and Tsallis C.} \REVIEW{The European Physical
  Journal B-Condensed Matter and Complex Systems}{1}{1998}{545}.

\bibitem{ferreira2015agreement}
\Name{Ferreira D.~S., Papa A.~R. \and Menezes R.} \REVIEW{Physics Letters
  A}{379}{2015}{669}.

\bibitem{blondel2008fast}
\Name{Blondel V.~D., Guillaume J.-L., Lambiotte R. \and Lefebvre E.} \REVIEW{J.
  Stat. Mech. Theor. Exp.}{10}{2008}{1000}.

\bibitem{barrat2004architecture}
\Name{Barrat A., Barthelemy M., Pastor-Satorras R. \and Vespignani A.}
  \REVIEW{Proceedings of the National Academy of Sciences of the United States
  of America}{101}{2004}{3747}.

\bibitem{brandes2001faster}
\Name{Brandes U.} \REVIEW{Journal of Mathematical Sociology}{25}{2001}{163}.

\end{thebibliography}

\end{document}